\documentstyle[12pt]{article}

\input epsf

\newcommand{\be}{\begin{equation}}
\newcommand{\ee}{\end{equation}}
\newcommand{\bea}{\begin{eqnarray}}
\newcommand{\eea}{\end{eqnarray}}
\newcommand{\ba}{\begin{array}}
\newcommand{\ea}{\end{array}}
\newcommand{\reseteq}{\setcounter{equation}{0}}

\newcommand{\hsp}[1]{\hspace*{#1pt}}
\newcommand{\hspf}{\; , \hspace*{5mm}}

\newcommand{\zr}[1]{\mbox{\hspace*{#1em}}}

\newcommand{\ZZ}{\mbox{\sf Z\zr{-0.45}Z}}

\newcommand{\EPL}[3]{Europhys.\ Lett.\ {\bf #1} \ (19#2) \ #3}

\newcommand{\IJMPB}[3]{Int.\ J.\ Mod.\ Phys.\ {\bf B#1} \ (19#2) \ #3}
\newcommand{\JMP}[3]{J.\ Math.\ Phys.\ {\bf #1} \ (19#2) \ #3}
\newcommand{\JPhysA}[3]{J.\ Phys.\ {\bf A#1} \ (19#2) \ #3}

\newcommand{\JSP}[3]{J.\ Stat.\ Phys.\ {\bf #1} \ (19#2) \ #3}

\newcommand{\MPL}[3]{Mod.\ Phys.\ Lett.\ {\bf #1} \ (19#2) \ #3}

\newcommand{\PhysicaA}[3]{Physica {\bf A#1} \ (19#2) \ #3}
\newcommand{\PhysRev}[3]{Phys.\ Rev.\ {\bf #1} \ (19#2) \ #3}

\newcommand{\PRSL}[3]{Proc.\ R.\ Soc.\ Lond.\ {\bf #1} \ (19#2) \ #3}

\newcommand{\RevMP}[3]{Rev.\ Math.\ Phys.\ {\bf #1} \ (19#2) \ #3}
\newcommand{\ZPhys}[3]{Z.\ Phys.\ {\bf #1} \ (19#2) \ #3}

\begin{document}


\begin{center}
{\LARGE\bf Partition Function Zeros \\[3mm]
 for Aperiodic Systems } \\[8mm]
\renewcommand{\thefootnote}{\fnsymbol{footnote}}
{\large\sc
  Michael Baake$^{1,2,}$\footnote[2]{Present address:
  Institut f\"{u}r Theoretische und Angewandte Physik,
  Universit\"{a}t Stuttgart, Pfaffenwaldring 57,
  70550 Stuttgart, Germany},
 \hsp{1} Uwe Grimm$^{1,3}$,
 \hsp{1} and \hsp{1} Carmelo Pisani$^{1,4}$} \\[4mm]
\renewcommand{\thefootnote}{\arabic{footnote}}
{\footnotesize \mbox{}\footnotemark[1]
 Department of Mathematics, University of Melbourne,\\[-2mm]
        Parkville, Victoria 3052, Australia} \\[2mm]
{\footnotesize \mbox{}\footnotemark[2]
 Institut f\"{u}r Theoretische Physik, Universit\"{a}t T\"{u}bingen,\\[-2mm]
        Auf der Morgenstelle 14, 72076 T\"{u}bingen, Germany} \\[2mm]
{\footnotesize \mbox{}\footnotemark[3]
 Instituut voor Theoretische Fysica, Universiteit van Amsterdam, \\[-2mm]
   Valckenierstraat 65, 1018 XE Amsterdam, The Netherlands} \\[2mm]
{\footnotesize \mbox{}\footnotemark[4]
 Department of Mathematics, La Trobe University, \\[-2mm]
   Bundoora, Victoria 3083, Australia} \\[6mm]
{\normalsize Preprint itap--11/93--2} \\[6mm]
\end{center}


\begin{quote}
{\small\sf
 The study of zeros of partition functions,
 initiated by Yang and Lee,
 provides an important qualitative and quantitative tool
 in the study of critical phenomena.
 This has frequently been used for periodic as well as
 hierarchical lattices. Here, we consider magnetic field and
 temperature zeros of Ising model partition functions
 on several aperiodic structures.
 In 1D, we analyze aperiodic chains obtained
 from substitution rules,
 the most prominent example being the Fibonacci chain. In 2D,
 we focus on the tenfold symmetric triangular tiling
 which allows efficient numerical treatment by means of
 corner transfer matrices.
}

\begin{center}
\noindent{\small\bf Key Words:} \hsp{2} \parbox[t]{116mm}{
\small Ising model, Lee-Yang zeros, non-periodic systems,
phase transitions, gap labelling}
\end{center}

\end{quote}



\renewcommand{\theequation}{\arabic{section}.\arabic{equation}}

%
%
%
%

\section{Introduction}
\reseteq

The study of critical phenomena by means of discrete spin systems, initiated
by Lenz and Ising \cite{Ising}, finally came of age by Onsager's spectacular
solution of the 2D field-free Ising model \cite{Onsager}.
He could show that the 2D Ising model
on the square lattice with ferromagnetic nearest neighbour interaction
exhibits an order-disorder phase transition
of second order at finite temperature
with the magnetization as order parameter.

The investigation of many other spin systems followed, as did the consideration
of Ising-type models on other lattices and graphs.
Some cases can still be solved
exactly, compare \cite{Baxter}, but this is an exception.
Consequently, one needs
complementary methods to tackle questions
like existence and location of critical
points and estimation of critical exponents, especially in higher dimensions.
One such technique is provided by the work of
Lee and Yang \cite{YangLee,LeeYang},
who investigated the distribution of
partition function zeros in the complex plane,
i.e., field variables or fugacity
were treated as a complex parameter.
Later, also the zeros in the complex temperature variable
were studied for various systems
\cite{Fisher,Jones,ItzykLuck,Carmelo}. These not only
provide information about the location of phase boundaries
but also about critical exponents, see e.g.~\cite{ItzykLuck}
where mainly hierarchical lattices are considered. For those,
the zero patterns form fractal structures known as Julia sets
whereas they are usually expected to lie on simple curves
for regular lattices, at least in the isotropic case.
It has been shown \cite{StephConz,vSaarKurtz} that for
anisotropic Ising models on two-dimensional regular lattices
the temperature zeros generically fill areas in the complex plane.

But what does ``simple curves'' mean? As we will show in a shortwhile,
already the Ising model on a modulated structure in 1D can result in
magnetic field zeros which {\em do} lie on a simple curve, but occupy
only a Cantor-like portion of it -- even in the thermodynamic limit!
Although the corresponding phenomenon in 2D does not typically show up
(or at least not that we could conclude so), the investigation of partition
function zeros for spin systems on non-periodic graphs with deflation/inflation
symmetry might fill the gap between the relatively well studied cases of
lattices and hierarchical graphs. This is the major motivation for the
present article, where we discuss the Ising model on quasiperiodic graphs
in 1D and 2D.

%
%
%
%

\section{Non-periodic 1D Ising model}
\reseteq

Let us consider a one-dimensional chain of
$N$ Ising spins \mbox{$\sigma_{j}\in\{\pm 1\}$},
\mbox{$j\!=\!1,2,\ldots,N$}, with periodic boundary conditions
(i.e., \mbox{$\sigma_{N+1}\!=\!\sigma_{1}$}).
The energy of a configuration
\mbox{$\sigma=(\sigma_{1},\sigma_{2},\ldots,\sigma_{N})$}
is given by
\be
E(\sigma) \;\; = \;\; -\,\sum_{j=1}^{N}\:
J_{j,j+1}\, \sigma_{j}\sigma_{j+1}
\; +\;  H_{j}\, \sigma_{j} \; .
\ee

Here, we concentrate on systems where the
coupling constants $J_{j,j+1}$ can
only take two different values
$J_{j,j+1}\in\{J_{a},J_{b}\}$ and where
the magnetic field is constant, i.e.\ $H_{j}\equiv H$.
The actual distribution of the two coupling constants
$J_{a}$ and $J_{b}$ along the chain is determined
by an infinite word in the letters $a$ and $b$ which
is obtained as the unique limit of certain two
letter substitution rules, see \cite{BGJ}.
In fact, we restrict ourselves to the
Fibonacci case which corresponds to the
substitution rule
\be
S:\;\; \ba{rcl}
 a & \rightarrow & b \\
 b & \rightarrow & ba \; . \ea
\ee
The length of the word $w_{n}=S(w_{n-1})$ obtained by
$n$ iterations from the initial word $w_{0}=a$ is
$f_{n+1}$, where the Fibonacci numbers $f_{n}$ are
defined by
\be
f_{0} \; = \; 0 \hspf
f_{1} \; = \; 1 \hspf
f_{n+1} \; = \; f_{n} + f_{n-1} \, .
\ee

Let us introduce the following notation
\be
K_{a} \; = \; \frac{J_{a}}{k_{B}T} \hspf
K_{b} \; = \; \frac{J_{b}}{k_{B}T} \hspf
h \; = \; \frac{H}{k_{B}T} \; ,
\ee
and
\be
z_{a}\; =\; \exp(2K_{a})\hspf
z_{b}\; =\; \exp(2K_{b})\hspf
w\; =\; \exp(2h) \; .
\ee
The two elementary transfer matrices $T_{a}$ and $T_{b}$
now read as follows
\be
T_{a,b} \;\; = \;\; (w \cdot z_{a,b})^{-1/2}\:
\left( \ba{cc}
 w \cdot z_{a,b} & \sqrt{w} \\ \sqrt{w} & z_{a,b}
\ea \right)
\;\; = \;\; (w \cdot z_{a,b})^{-1/2}\:\tilde{T}_{a,b}
\ee
and, in general, do {\em not} commute with each other.

The recursive definition of the sequence gives
rise to a recurrence formula for the transfer matrices
\be
T_{0}\;\; =\;\; T_{a}\hspf
T_{1}\;\; =\;\; T_{b}\hspf
T_{n+1}\;\; = \;\; T_{n}\cdot T_{n-1} \; ,
\ee
where $T_{n}$ denotes the transfer matrix of the
chain that corresponds to the $n$-th iteration step.
Hence the partition function
$Z_{n}(z_{a},z_{b},w)=\mbox{tr}(T_{n})$ is
essentially (i.e., up to an overall factor)
a polynomial in its three variables.
It is also
possible to write down a recurrence relation
for the partition function itself,
see \cite{BGJ} for details.
In this way, it is really easy to generate the
partition function for very large systems exactly (e.g., by means
of algebraic manipulation packages).
Our problem at hand has thus been reduced to the
task of computing the roots of a polynomial.

Let us have a look at the pattern of zeros of
the partition functions $Z_{n}(z_{a},z_{b},w)$
in the field variable $w$.
If both couplings are ferromagnetic
(\mbox{$z_{a}\geq 1$}, \mbox{$z_{b}\geq 1$}),
the zeros lie on the unit circle. For purely
anti-ferromagnetic coupling,
(\mbox{$z_{a}\leq 1$}, \mbox{$z_{b}\leq 1$}),
they are on the imaginary axis whereas the
``mixed'' case turns out to be complicated and
will not be discussed here.

For simplicity, we will, from now on, stick to the
purely ferromagnetic regime. If the chain were periodic,
the zeros would be distributed homogeneously on the unit circle
(they can be calculated analytically).
In the thermodynamic limit, they fill the circle densely except
near the point (1,0) on the real axis, where we are left with
a gap since we do not have a phase transition at finite temperature
(i.e., for $0 < T < \infty$). One might perhaps expect the very same
situation in the Fibonacci case, but the latter is always good for a surprise.

In Fig.~1, we show the location of the
zeros of $Z_{n}(z_{a},z_{b},w)$
in the complex $w$-plane for ferromagnetic
couplings $z_{a}=3/2$, $z_{b}=100$
(which corresponds to $K_{a}\approx 0.203$ and
$K_{b}\approx 2.306$)
and $n=8,9,10$.
As expected, the zeros are
clearly located on the unit circle, and there is still a large gap
near (1,0) as it must be because we still cannot have a phase transition
at finite temperature.
However, an {\em additional}\/ gap structure
in the distribution of the zeros on the unit circle is
apparent. It turns out that this gap structure does not
depend on the actual values of the coupling constants
(as long as they are still ferromagnetic),
just the gap widths change and the gaps vanish if $z_{a}$
and $z_{b}$ become equal which of course corresponds to
the periodic case. The large difference between the two
couplings used in our pictures was chosen to show the effect
clearly -- for a small difference one might miss it, though
it is still there!

\vspace{5mm}
\begin{footnotesize}
\begin{center}
{\bf Figure~1:}\hsp{2}
\parbox[t]{120mm}{
Zeros of the partition function
$Z_{n}(z_{a},z_{b},w)$ of the
Fibonacci Ising chain in the field variable $w=\exp(2h)$
for $z_{a}=3/2$, $z_{b}=100$, and $n=8,9,10$.}\\*[5mm]
\epsfxsize=\textwidth \epsfbox[0 94 288 194]{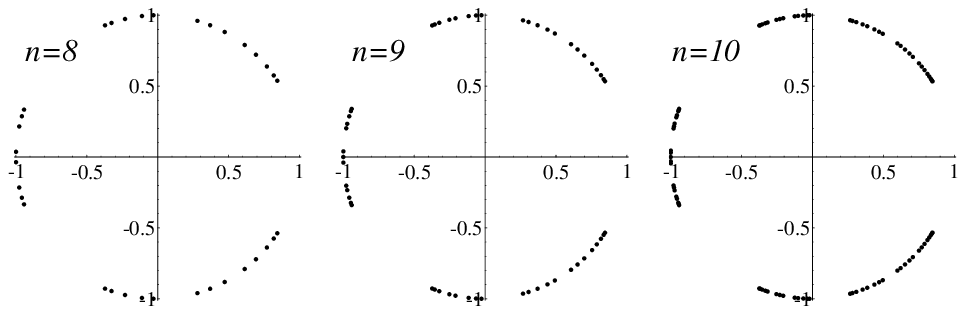}
\end{center}
\end{footnotesize}
\vspace{5mm}

In Fig.~2, we present the integrated density of the zeros
on the unit circle (i.e., the integrated density of their
angles (in units of $2\pi$) with respect to the real axis)
for the same set of parameter values as in Fig.~1.
It turns out that this has exactly the structure which
one would expect from the general gap labeling
theorem of Bellissard and
coworkers \cite{Belli86,BelliBovGhez92,BGJ}
which however applies to
the integrated density of states (IDOS) of
energy spectra of certain discrete Hamiltonians. There, the limit set of
the IDOS values at the gaps has the form
(see \cite{BGJ} and references therein)
\be
{\cal G} \;\; = \;\;
\left\{\;\frac{\mu}{\tau} + \frac{\nu}{\tau^{2}}
\;\left|\;\rule{0mm}{4mm}
\mu, \nu\in\ZZ\;\right.\right\} \;\cap\;
\left[\, 0 , 1\, \right]
\ee
for the Fibonacci case, where $\tau = \frac{1 + \sqrt{5}}{2}$ is the golden
ratio.
The widest gaps in Fig.~2 have been labeled with the
corresponding indices $(\mu,\nu)$ which for the finite
system belong to the rational values
\be
(\mu,\nu) \, : \;\;
\frac{\mu f_{n} + \nu f_{n-1}}{f_{n+1}}
\;\;\;
\stackrel{n \rightarrow \infty}{\longrightarrow}
\;\;\;
\frac{\mu}{\tau} + \frac{\nu}{\tau^{2}}
\label{finitegap}
\ee
of the integrated density. Note that any two successive
Fibonacci numbers $f_{n-1}$ and $f_{n}$ are
coprime, hence every integer can be written as a
linear combination of them with integral coefficients.
Generically, it appears that at all the ``allowed''
values one actually
observes gaps in the density distribution,
i.e., ``all gaps are open''.
The obvious symmetry
in the pictures which relates the gaps with labels $(\mu,\nu)$ and
$(1-\mu,1-\nu)$ stems from the reflection symmetry
of the zero pattern with respect to the real axis which
corresponds to a change of the sign in the magnetic field $h$.

\vspace{5mm}
\begin{footnotesize}
\begin{center}
{\bf Figure~2:}\hsp{2}
\parbox[t]{120mm}{
Integrated density of the zeros in Fig.~1
on the unit circle. The corresponding gap labels
(see Eq.~(\ref{finitegap})) for the widest gaps are also shown.}\\*[5mm]
\epsfxsize=\textwidth \epsfbox[0 94 288 194]{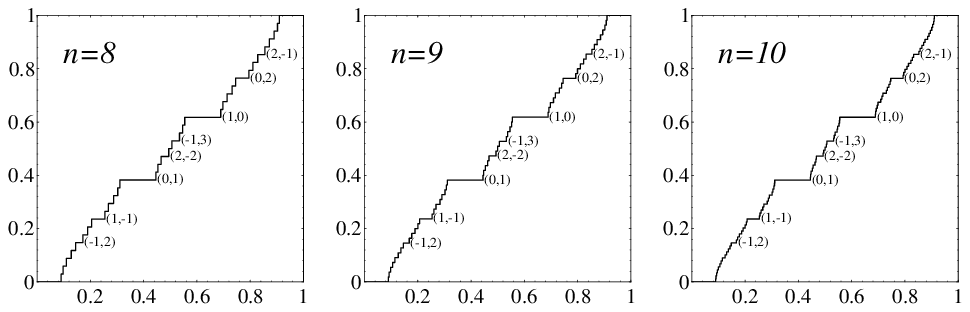}
\end{center}
\end{footnotesize}
\vspace{5mm}

Admittedly, we did not present a rigorous argument which explains
why the gap labeling theorem for the IDOS of one-dimensional
Sch\"{o}dinger operators \cite{Belli86,BelliBovGhez92,BGJ}
describes the distributions of our partition function zeros
on the unit circle. However, the agreement is certainly
convincing and suggests to think of the zeros
(or rather of their arguments) as the eigenvalues of a
Schr\"{o}dinger type operator. This is also one relatively easy
way to prove the Lee-Yang circle theorem (or the corresponding
``line theorem'' in the antiferromagnetic case) for the periodic 1D
Ising model. In this context, we also mention an old idea of Hilbert's,
namely the possible connection between the imaginary parts of
the (non-trivial) zeros of the Riemann $\zeta$-function and
the eigenvalues of a -- so far unknown --
hermitian Hamiltonian. This has recently also
been linked to some typical aspects of
quantum chaos, see \cite{Berry} and references therein.

%
%
%
%

\section{Ising model on the triangle tiling}
\reseteq

The 1D case was presented for two main reasons.
On the one hand, calculations
are either possible analytically or can be made rigorous.
On the other hand,
a new phenomenon -- the appearance of gaps -- could be seen.

Nevertheless, investigations of Ising models
on graphs of higher dimension
are in order. If one is not interested in
approximative calculations (which we are not),
one encounters true difficulties on the non-periodic ground.
Even the existence of local deflation/inflation symmetry
does not seem to allow exact renormalization schemes
for electronic models, compare \cite{Sire90,You92},
and we were not able so far to find one for Ising type
models, either.

So, the best thing one can then do is
the exact calculation of the partition
sum for finite patches, followed by a numerical approach
of the thermodynamic limit.
Even this is a difficult task which requires
a good choice of the graph,
i.e., the non-periodic tiling.
{}From our experience with Ising quantum chains
in 1D \cite{GB93} -- which can be seen as anisotropic
limits of 2D classical systems --
we decided to take a quasiperiodic example
in order to stay free of effects
caused by uncontrollable fluctuations.

\vspace{5mm}
\begin{footnotesize}
\begin{center}
{\bf Figure~3:}\hsp{2}
\parbox[t]{120mm}{
Initial decagonal patch of the T\"{u}bingen
triangle pattern.}\\*[5mm]
\epsfysize=150pt \epsfbox[-300 0 588 288]{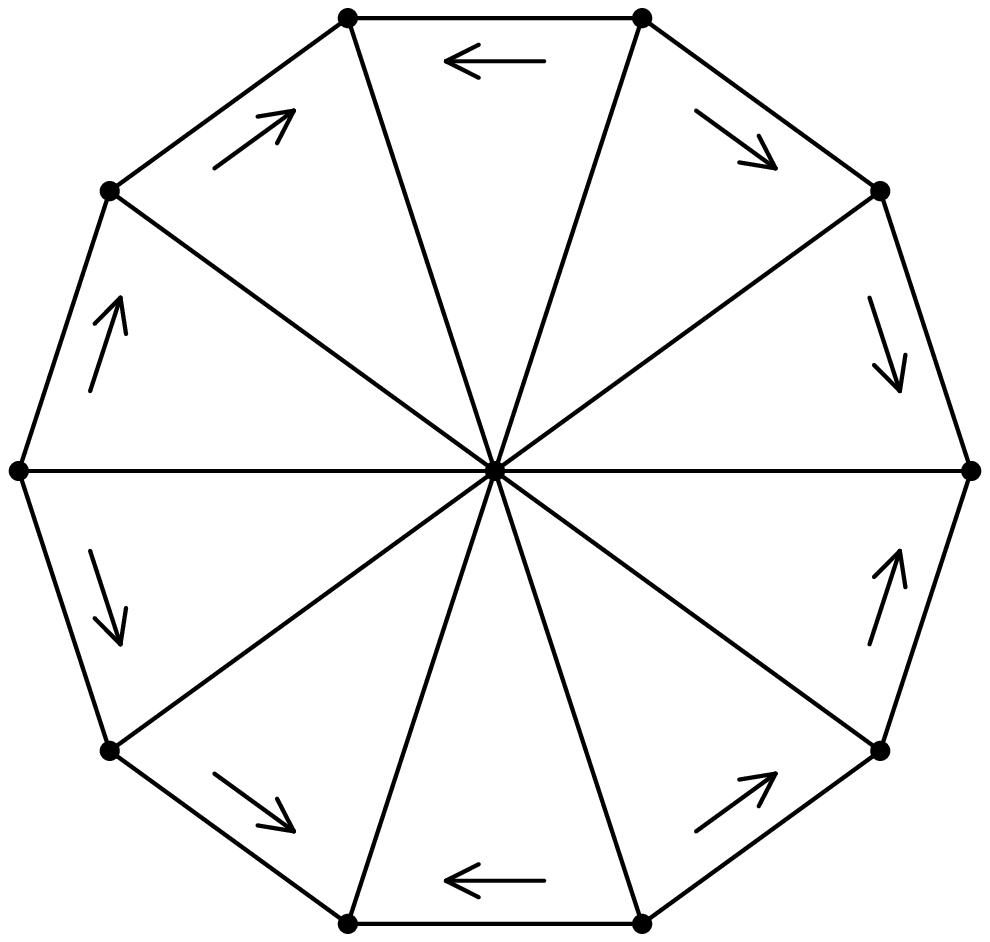}
\end{center}
\end{footnotesize}
\vspace{5mm}

The triangle pattern developed in T\"ubingen
\cite{triangle1,triangle2,triangle3}
has the advantage that we can start from a
patch with decagonal boundary that
is -- in a natural way -- partitioned into {\em sectors},
each covering an angle
of \mbox{$2 \pi / 10$}, compare Fig.~3.
This structure suggests the use
of corner transfer matrices,
compare \cite{Baxter}, to calculate the partition sum
and the magnetization
at the centre of the patch. However,
the patch is not a repetition of ten
equal sectors, wherefore we cannot reduce
the problem to the CTM of one
single sector. On the other hand, the CTM's of sectors of opposite
orientation (indicated by the arrow in the inital patch) are the transposed
matrices of each other. Giving those spins half weight
which lie on the boundaries of
the sectors, we can obtain the partition function as
\be \label{ctm}
   Z \;\; =\;\;  \mbox{tr}\, \left(
(M^2 M^t M^2) \cdot (M^2 M^t M^2)^t \right) \; .
\ee
This can then be evaluated again by algebraic manipulation packages.

In what follows, we use the same notation as in the
Fibonacci case above but with indices $s$ (for short)
and $\ell$ (for long) in place of $a$ and $b$.
In particular, we also assume that the magnetic
field is uniform.
Here, $Z_{n}(z_{s},z_{\ell},w)$ now
denotes the partition function of
the Ising model on the patch which is obtained by
$n$ deflation steps (see Fig.~4) from the initial
decagonal patch shown in Fig.~3 (which corresponds to $n=0$).
It is impossible to define periodic boundary
conditions on our tiling without destroying the tenfold symmetry
and the sector structure, wherefore we either
use fixed boundary conditions (i.e., all spins on the
decagonal boundary of the patch are frozen to be $+1$)
or free boundary conditions (i.e.,
we sum over all values of the boundary spins).

\vspace{5mm}
\begin{footnotesize}
\begin{center}
{\bf Figure~4:}\hsp{2}
\parbox[t]{120mm}{
Deflation rules for the T\"{u}bingen triangle
pattern.}\\*[5mm]
\epsfysize=150pt \epsfbox[-150 60 488 248]{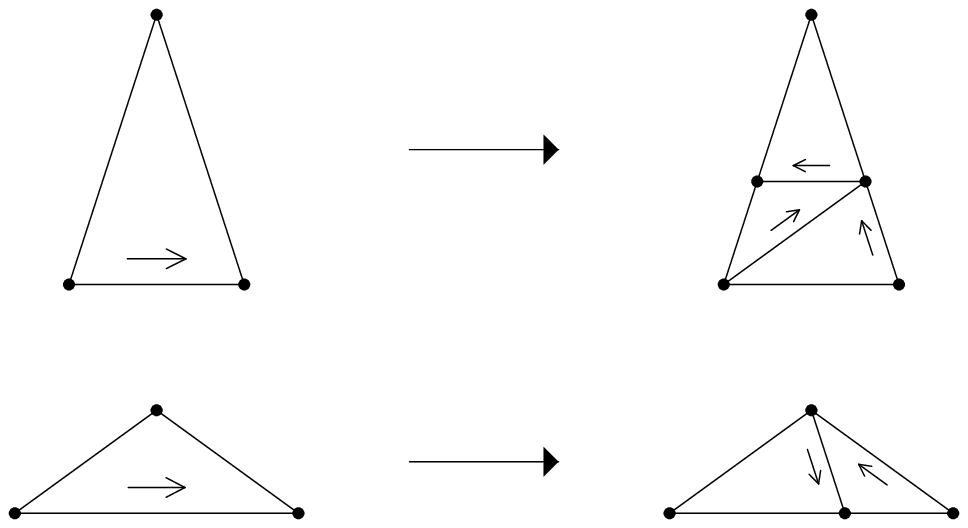}
\end{center}
\end{footnotesize}

\begin{footnotesize}
\begin{center}
{\bf Figure~5:}\hsp{2}
\parbox[t]{120mm}{
One sector of Fig.~3 after $n$ deflation steps
(see Fig.~4) for $n\leq 4$. The dashed line indicates
a sector that also leads to a patch with decagonal boundary,
which however is not obtained by entire deflation steps from
the initial patch shown in Fig.~3.}\\*[5mm]
\epsfbox[-100 0 388 288]{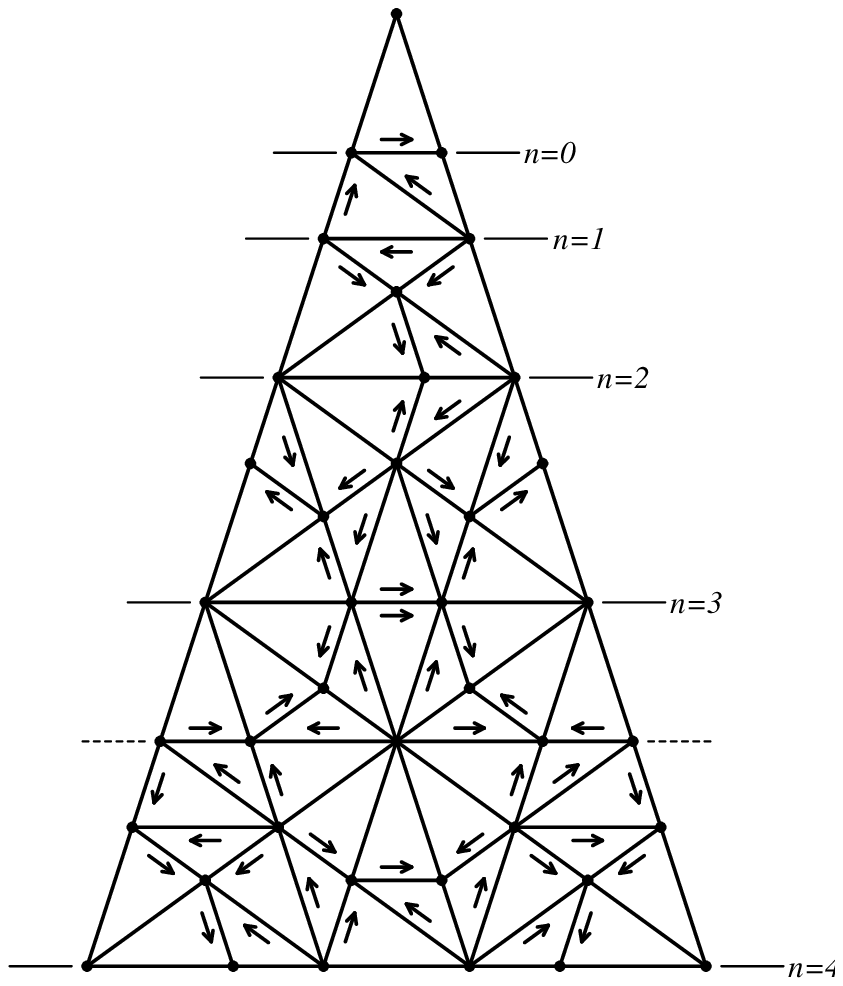}
\end{center}
\end{footnotesize}
\vspace{5mm}

For fixed boundary conditions, the zeros of
the partition function $Z_{n}(z_{s},z_{\ell},w)$
(for finite $n$) in the field variable $w$ are no
longer located on the unit circle,
as this case (in contrast to the free boundary case) is not
covered by the circle theorem of Lee and Yang \cite{LeeYang}.
This is clearly seen in Fig.~6 which
shows the locations of the zeros in the
complex $w$-plane for couplings $z_{s}=z_{\ell}=4$
and three different patch sizes.
It is plausible that the zeros approach the unit circle
in the thermodynamic limit.
However, the angular distribution of the zeros is remarkably regular,
there is no apparent gap structure as in the one-dimensional
case. We also computed the zeros for different
couplings $z_{s}$ and $z_{\ell}$ for fixed and free
boundary conditions which show the same behaviour.

\vspace{5mm}
\begin{footnotesize}
\begin{center}
{\bf Figure~6:}\hsp{2}
\parbox[t]{120mm}{
Zeros of the partition function $Z_{n}(4,4,w)$
(fixed boundary conditions)
in the field variable $w$ for $n=1,2,3$.}\\*[5mm]
\epsfbox[-100 0 388 288]{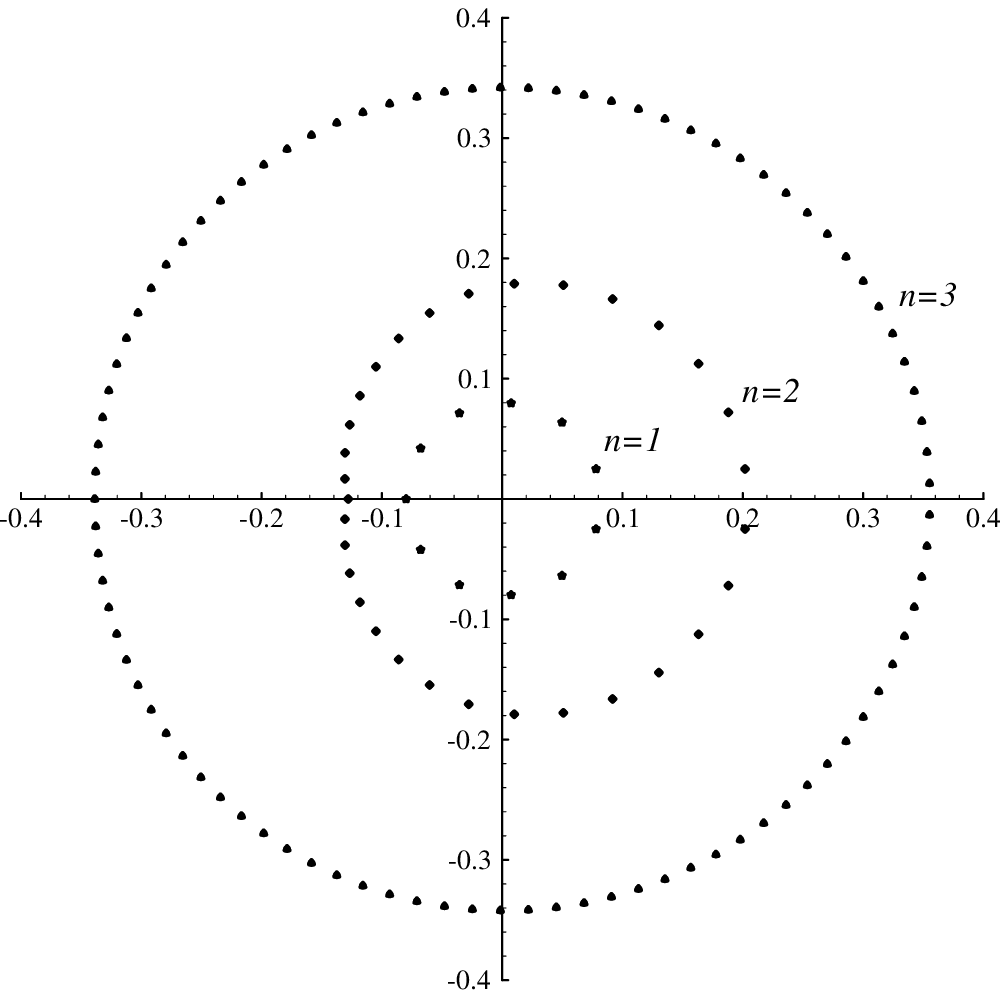}
\end{center}
\end{footnotesize}
\vspace{5mm}

Let us also have a look at the zeros in the other variables.
Here, we restrict ourselves to the zero-field
partition function $Z_{n}(z_{s},z_{\ell},1)$
and use fixed boundary conditions.
For the ``isotropic'' case $z=z_{s}=z_{\ell}$, the zeros
of $Z_{3}(z,z,1)$ in the variable $z$ are shown in
Fig.~7. Although the distribution of the zeros in the
complex $z$ plane is apparently not simple
(i.e., the zeros do not appear to lie on simple
curves), the zeros close to the real axis contain
information about the location of the critical point
and, in principle, also about the critical exponents
\cite{ItzykLuck}. In Table~1, we give the numerical
values of the zero closest to the real line
with a real part greater than one for the following five choices
of $(z_{s},z_{\ell})$: $(1,z)$, $(z,z^2)$, $(z,z)$, $(z^2,z)$,
and $(z,1)$,
i.e., in addition to the ``isotropic'' case we look at those cases where
one of the coupling constants vanishes and where one coupling constant is twice
as large as the other.

\vspace{5mm}
\begin{footnotesize}
\begin{center}
{\bf Figure~7:}\hsp{2}
\parbox[t]{120mm}{
Zeros of the partition function $Z_{3}(z,z,1)$
(fixed boundary conditions)
for the ``isotropic'' zero-field case (equal coupling for long and
short bonds) in the temperature variable $z$.}\\*[5mm]
\epsfbox[-100 0 388 288]{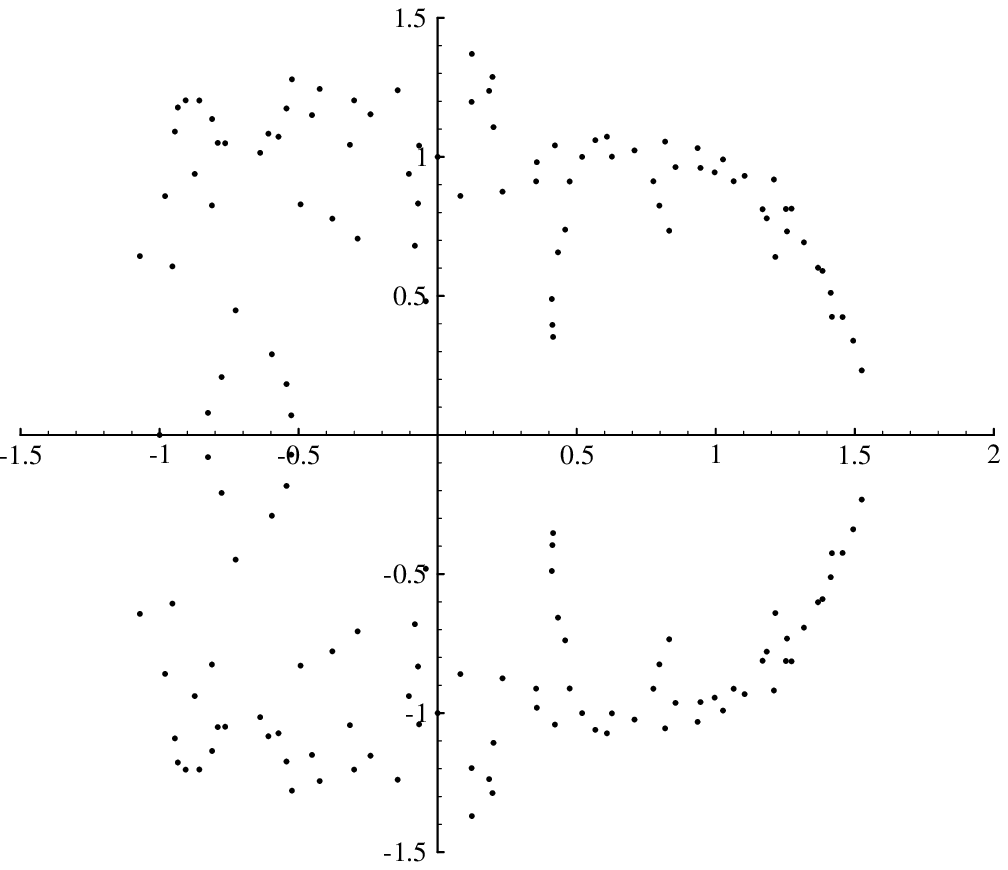}
\end{center}
\end{footnotesize}
\vspace{5mm}

\vspace{5mm}
\noindent\parbox{\textwidth}{
\begin{footnotesize}
\begin{center}
{\bf Table 1:}\hsp{2}
\parbox[t]{120mm}{Partition function zero located
closest to the real line
and with a real part greater than one, for
vanishing magnetic field and fixed boundary conditions.}\\*[5mm]
\begin{tabular}{|c|c|c|c|c|c|}
\hline
$n$ & $z_{s}=1$ & $z_{s}=z_{\ell}^2$ &
$z_{s}=z_{\ell}$ & $z_{s}^2=z_{\ell}$ & $z_{\ell}=1$ \\ \hline
1 & $1.3514\pm 0.9560 i$ & $1.2396\pm 0.2819 i$ &
    $1.3249\pm 0.4315 i$ & $1.1987\pm 0.2492 i$ & $1.4511\pm 1.3404 i$ \\
2 & $1.8056\pm 0.8702 i$ & $1.3270\pm 0.2006 i$ &
    $1.4709\pm 0.3238 i$ & $1.2788\pm 0.1844 i$ & $2.4447\pm 1.3460 i$ \\
3 & $1.9089\pm 0.6472 i$ & $1.3670\pm 0.1412 i$ &
    $1.5249\pm 0.2322 i$ & $1.2983\pm 0.1320 i$ & $3.0508\pm 1.1471 i$ \\
\hline
\end{tabular}
\end{center}
\end{footnotesize} }
\vspace{5mm}

The approximants for the critical couplings are in good
agreement with our values obtained from the behaviour of the
specific heat and the centre spin magnetization of the model
(with the exception of $z_{\ell}=1$ where the centre spin is isolated),
but we omit details here. Let us only remark that the nature of the critical
point looks very much like that of the periodic case though further
calculations are needed to approve that.

%
%
%

\section{Concluding Remarks}
\reseteq

The distribution of partition function zeros on the unit circle shows
a gap structure for ferromagnetic Ising models on aperiodic chains.
Although we have only demonstrated this phenomenon for the ubiquitious
Fibonacci chain, one can translate each step to {\em any}\/ of the other
chains obtained by substitution rules -- no matter whether one restricts
oneself to the two-letter case or not.

The 2D case did not show any apparent gap structure
for the magnetic field zeros (and thus resembles the
situation of the electronic spectra again).
The partition function zeros in the temperature variable
do not seem to lie on simple curves,
even in the ``isotropic'' case where the couplings for short and
long bonds are identical -- more work is to be done to clarify
this point.
On the other hand, one can clearly see the existence of a phase transition
because the zeros ``pinch'' the real axis. For given (finite) coupling
constants,
the critical point has finite $T_c$.

Now, looking at the limiting cases $z_s=1$ or $z_{\ell}=1$,
one gets the impression
that $T_c$ heads to 0, quite similarly as in the case of the square lattice
\cite{Baxter}. However, the corresponding graph
with one type of bonds ``switched off''
is by no means one-dimensional (or at least not in an obvious way) --
wherefore the question arises what this means. To clear this up, one should
treat first a model where the location of the critical point can be found
exactly, without numerical estimates. This is indeed possible for another
quasiperiodic pattern, the so-called Labyrinth tiling. The same phenomenom
occurs there, and we will discuss details elsewhere \cite{Laby}.

%
%
%
%

\section*{Acknowledgements}

This work was supported by the Australian Research Council,
Deutsche Forschungsgemeinschaft and
the EC Human Capital and Mobility Program.



\end{document}